\documentclass[aps,prb,floatfix,preview,citeautoscript,longbibliography,superscriptaddress]{revtex4}
\usepackage{graphicx}
\usepackage{ulem}
\usepackage{bm,amsmath,amssymb,mathrsfs}
\usepackage[colorlinks=true, linkcolor=blue, citecolor=blue, urlcolor=blue, bookmarks=false]{hyperref} 
\usepackage[usenames]{xcolor}
\hypersetup{pdfborder=0 0 0,colorlinks=true,citecolor=blue,linkcolor=blue}
\usepackage[final]{pdfpages}

\begin{document}
\title{Magnon-phonon relaxation in yttrium iron garnet from first principles}
\author{Yi Liu}
\author{Li-Shan Xie}
\author{Zhe Yuan}
\affiliation{The Center for Advanced Quantum Studies and Department of Physics, Beijing Normal University, 100875 Beijing, China}
\author{Ke Xia}
\affiliation{The Center for Advanced Quantum Studies and Department of Physics, Beijing Normal University, 100875 Beijing, China}
\affiliation{Synergetic Innovation Center for Quantum Effects and Applications (SICQEA), Hunan Normal University, Changsha 410081, China}
\date{\today}
\begin{abstract}
We combine the theoretical method of calculating spin wave excitation with the finite-temperature modeling and calculate the magnon-phonon relaxation time in the technologically important material Yttrium iron garnet (YIG) from first principles. The finite lifetime of magnon excitation is found to arise from the fluctuation of the exchange interaction of magnetic atoms in YIG. At room temperature, the magnon spectra have significant broadening that is used to extract the magnon-phonon relaxation time quantitatively.  The latter is a phenomenological parameter of great importance in YIG-based spintronics research. We find that the magnon-phonon relaxation time for the optical magnon is a constant while that for the acoustic magnon is proportional to $1/k^2$ in the long-wavelength regime. 
\end{abstract}
\maketitle

Yttrium iron garnet (Y$_3$Fe$_5$O$_{12}$, YIG), a ferrimagnetic insulator, has been extensively applied in the spintronics experiments, such as the investigations of the spin Seebeck effect \cite{Uchida:natm10}, spin Hall magnetoresistance \cite{Nakayama:prl13} and cavity magnon polariton \cite{Bai:prl15}. In particular, YIG is an ideal magnetic material for transport study of a pure spin current because electronic transport can be completely eliminated. For example, a spin wave is allowed to propagate in YIG over a long distance \cite{Cornelissen:natp10,Brandon:prb15} due to its ultralow Gilbert damping\cite{Sun:apl12} of YIG. Unlike the ferromagnetic metals, where the Gilbert damping is dominated by the conduction electrons near the Fermi level \cite{Kambersky:cjp76, Gilmore:prl07}, spin-lattice interaction is the main mechanism of dissipating angular momentum and magnetic energy during the magnetization dynamics of YIG. Fundamental research on the spin-lattice interaction can be traced back to the early works in 1950s, when Abrahams and Kittel developed a phenomenological theory of the magnetoelastic effects in magnetic metals.\cite{Abrahams:pr52,Kittel:rmp53,Kittel:pr58} They introduced a phenomenological relaxation time to characterized the spin-lattice interaction. Later Sanders and Walton performed an experimental measurement investigating the magnon-phonon relaxation time $\tau_{\textrm mp}$ in magnetic insulators.\cite{Sanders:prb77} Recently, $\tau_{\textrm mp}$ is found to be crucially important to describe many YIG-based spin transport phenomena \cite{Vittoria:prb10,Rueckriegel:prb14,Cornelissen:prb16}. Nevertheless, a quantitative estimation of the phenomenological relaxation time in a complex magnetic material like YIG is essentially nontrivial; $\tau_{\textrm mp}$ can be expressed in terms of the magnetoelastic coupling constants, which are, however, still short of reliable evaluation experimentally or theoretically. 

The present work aims at a quantitative evaluation of the magnon-phonon relaxation time in YIG using a physically transparent and numerically reliable method. More specifically, we combine the first-principles method of calculating magnon spectrum \cite{Xie:prb17} and a finite-temperature modeling by displacing atoms from their equilibrium positions \cite{Liu:prb11,Liu:prb15}, so as to obtain the fluctuation of the exchange interaction induced by lattice vibration. Such a computational scheme allows us to determine the magnon excitation mediated by phonons without imposing any phenomenological parameters. As a consequence, the fluctuation of the exchange interaction results in a broadening of the magnon spectrum. Then we are able to extract the magnon-phonon relaxation time based on this broadening. 

\begin{figure}[b]
\begin{center}
\includegraphics[width=0.9\columnwidth]{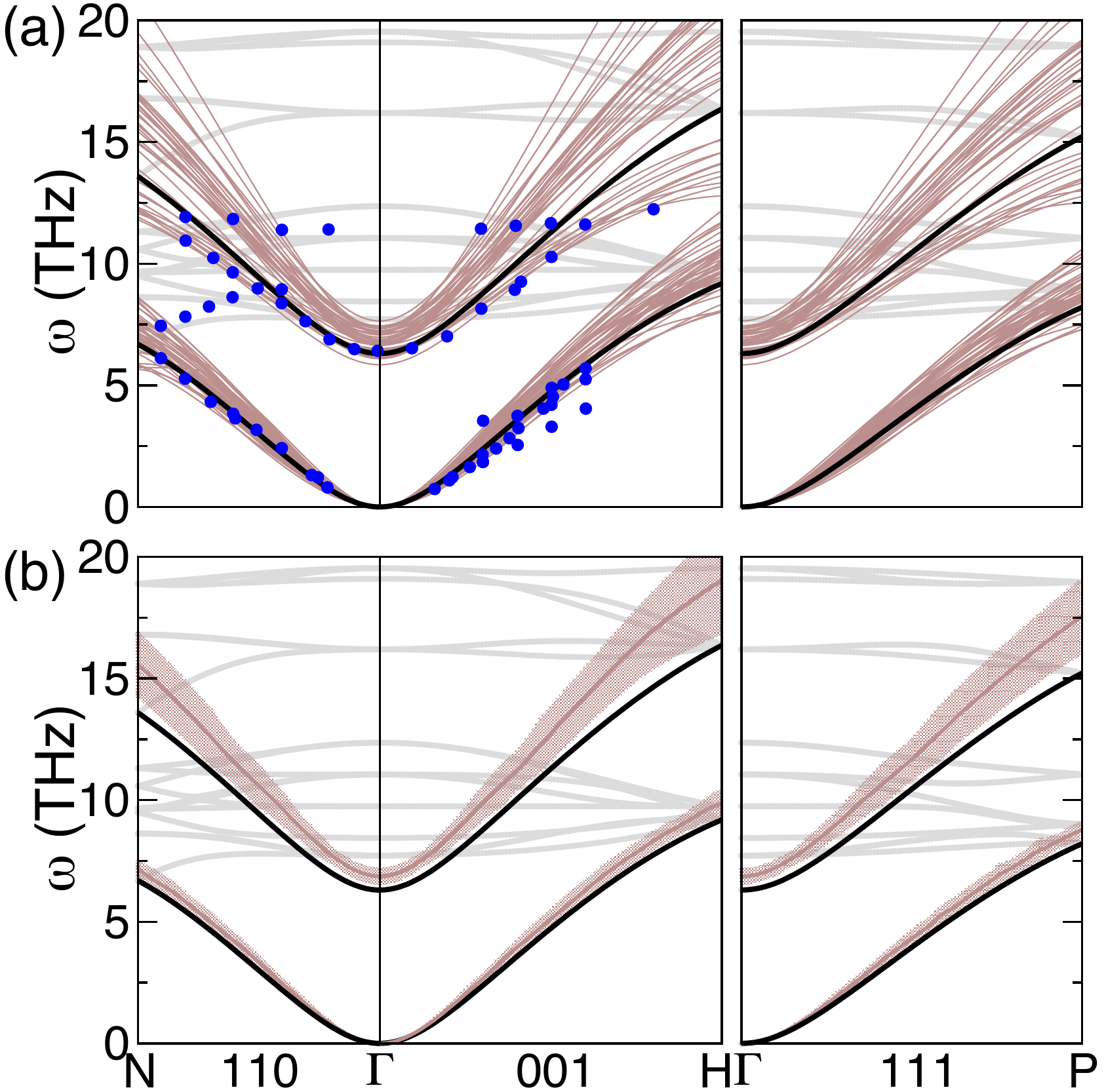}
\end{center}
\caption{(a) Calculated magnon spectra of YIG. At zero temperature, the dispersion of the acoustic magnon mode and the lowest-energy optical mode are shown as the black curves, while the other modes are plotted as grey curves. The experimental data (blue dots) are shown for comparison \cite{Plant:jpc77,Plant:jpc83}. The brown curves correspond to the calculated acoustic and optical magnon modes using 40 random phonon configurations at room temperature. (b) The same as (a) but the calculated acoustic and optical magnon spectra are replotted a spectrum with errorbars (shadows) based on the 40 brown curves in (a).}\label{fig:spectrum}
\end{figure}
The electronic structure of the bulk YIG is calculated self-consistently based upon the generalized gradient approximation of the density functional theory, which has been implemented in the Vienna \textit{ab initio} simulation package (VASP)\cite{Kresse:prb93,Kresse:prb96}. We choose $U-J=2.7$~eV in all the calculations to account for the strongly correlated electronic states. The experimental lattice constant 12.375~{\AA} is used for the body-centered cubic (bcc) unit cell, which contains 80 atoms in total. At the ground state, a Fe atom on one of the twelve $d$ sites in the unit cell has a magnetic moment aligned oppositely with the moment of a Fe atom on one of the eight $a$ sites, whose magnitudes are both approximately $5~\mu_B$. The other atoms are not magnetized. By flipping one magnetic moment of the Fe atoms, we are able to obtain the total energy of the corresponding metastable state. Comparing the energy difference between the metastable states and the ground state, we can calculate the exchange constants between a pair of Fe atoms both on the $d$ or $a$ sites ($J_{dd}$ and $J_{aa}$), or on two different sites ($J_{ad}$). The computational details to determine the exchange constants can be found in Ref.~\onlinecite{Xie:prb17}. Having obtained the exchange constants, we determine the magnon spectrum using the ``frozen magnon method'' \cite{Halilov:epl97,Halilov:prb98}. As an example, the magnon spectra of the perfectly crystalline YIG are shown by the black and grey curves in Fig.~\ref{fig:spectrum}(a) that are in good agreement with the experimental measurement (blue dots).\cite{Plant:jpc77,Plant:jpc83}

Note that these (black and grey) magnon spectra in Fig.~\ref{fig:spectrum}(a) are obtained without any phonons corresponding to the case at zero temperature. To examine the influence of phonons, we apply the recently developed computational scheme to account for the temperature-induced atomic vibration \cite{Liu:prb11,Liu:prb15}, i.e., at finite temperature, atoms in the bulk YIG displaced away from their equilibrium positions in a crystalline lattice. The magnitude of the displacements increases with increasing temperature. In the practical calculation, the displacement $\mathbf{u}_i$ of the $i$-th atom with its mass $M_i$ follows a random Gaussian distribution and we have the statistical mean square of the displacements determined by the Debye model,
\begin{equation}
\langle|\mathbf u_i|^2\rangle=\frac{9\hbar^2}{M_ik_\textrm{B}}\Theta_\textrm{D}\left(\frac{T^2}{\Theta_\textrm{D}^2}\int^\frac{\Theta_{\rm D}}{T}_0\frac{x}{e^x-1}dx+\frac{1}{4} \right).
\end{equation}
In this paper, we focus on the properties at room temperature, $T=300~K$ and the Debye temperature $\Theta=403$~K is chosen from experiment\cite{Modi:jms05}. It is worth noting that we do not take the temperature-induced magnon-magnon interaction into account in our calculation, which has been investigated with atomistic spin dynamics simulation \cite{Barker:prl16}.

Strictly speaking, the exchange interaction of each pair of magnetic Fe atoms in the disordered YIG depends on the specific distance between them. So there may be different numerical values of $J_{ad}$, $J_{dd}$, or $J_{aa}$  in a unit cell of YIG with lattice disorder. When applying the method in Ref.~\onlinecite{Xie:prb17} to determine the $J$'s, we do not distinguish the distance-dependent exchange interaction, but instead average over pairs of Fe atoms in one disordered configuration to obtain an effective value of $\bar{J}_{ad}$, $\bar J_{dd}$, and $\bar J_{aa}$. Using the effective exchange constants, we calculate the corresponding magnon spectrum for this disordered configuration. In practice, we consider 40 different disordered configurations and the resulting acoustic and the lowest-energy optical magnon dispersions are plotted by the brown curves in Fig.~\ref{fig:spectrum}(a). The other optical branches of magnons obtained at room temperature are now not shown for simplicity. The brown curves basically superimpose on the zero-temperature spectra curves (the black ones) and show significant spread in energy.
 
The spread of the magnon dispersion can be interpreted in terms of a spectral function $A({\bf{k}},\omega)$, which defines the probability density of the magnon mode having the wavevector ${\bf{k}}$ and the frequency $\omega$. Following a standard expression using the retarded Green's function, \cite{Doniach:98} the spectral function has the form of a Lorentzian function:
\begin{equation}
A({\bf{k}},\omega)\propto\sum_i\frac{\gamma_{i,\bf{k}}}{(\omega-\tilde{\omega}_{i,\bf{k}})^2+\gamma^2_{i,\bf{k}}},
\end{equation}
where $\tilde{\omega}_{i,\bf{k}}$ is the effective (median) frequency of the $i$-th branch magnon and $\gamma_{i,\bf{k}}$ is the width of this mode. In the case of our 40 different configurations, $\omega_{i,{\bf{k}}}^n$ with $n=1,...,40$, we determine $\tilde{\omega}_{i,\bf{k}}$ at each ${\bf k}$ point and the width, respectively, by
\begin{equation}
\tilde{\omega}_{i,\bf{k}}=\mathrm{median}(\omega^n_{i,{\bf{k}}}),
\end{equation}
and
\begin{equation}
\gamma_{i,\bf{k}}=\mathrm{median}(\vert{\omega}^n_{i,{\bf{k}}}-\tilde{\omega}_{i,\bf{k}}\vert).
\end{equation}
A careful numerical test shows that $\tilde{\omega}_{i,\bf{k}}$ and $\gamma_{i,\bf{k}}$ for the acoustic and the lowest-energy optical magnons are both well converged with 40 configurations. We then plot $\tilde{\omega}_{i,\bf{k}}$ and $\gamma_{i,\bf{k}}$ of the calculated room-temperature spectra in Fig.~\ref{fig:spectrum}(b) as the brown curves with error bars. The effective magnon frequency $\tilde{\omega}_{i,\bf{k}}$ agrees globally with the zero temperature spectra despite of a slight blue shift, which is discussed later. Such global agreement indicates that the exchange interaction between the Fe atoms in YIG does not change much due to the temperature-induced ionic vibration, and justifies our computational framework of applying the frozen lattice disorder. 

\begin{figure}[t]
\begin{center}
\includegraphics[width=0.9\columnwidth]{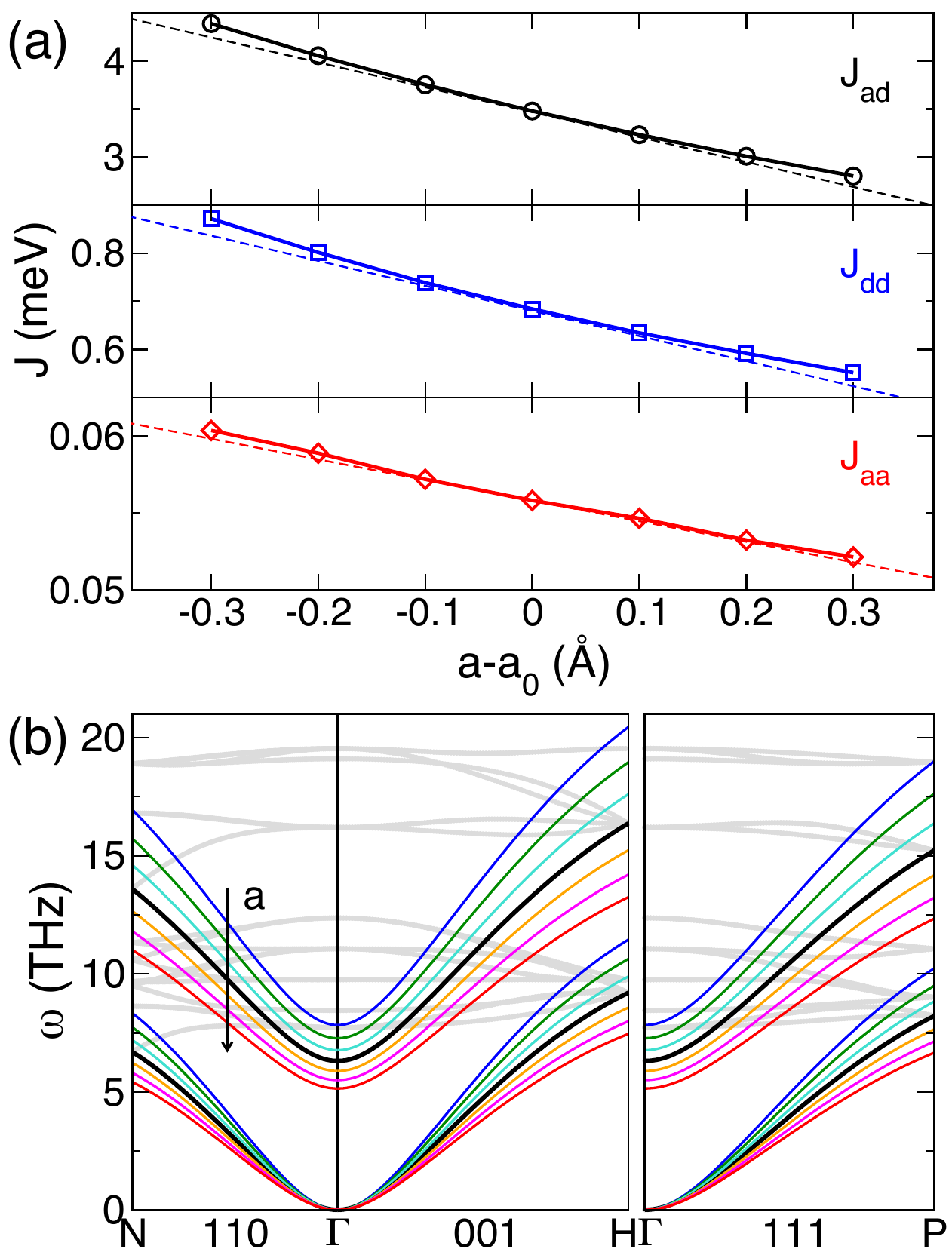}
\caption{(a) Calculated exchange interaction in YIG as a function of the lattice constant. The dashed lines illustrate the linear dependence. (b) Calculated spectra of the acoustic and the lowest-energy optical magnons of YIG as a function of the lattice constant. The black grey lines are the same as in Fig.~\ref{fig:spectrum}. The spectra of different colors corresponds to artificially changed lattice constants in (a). Thermal lattice disorder is not included in these calculations.}\label{fig:change_a}
\end{center}
\end{figure} 
The magnon frequency at room temperature in Fig.~\ref{fig:spectrum} is slightly higher than that at zero temperature. The blue shift is larger for the optical mode and at the edge of the Brillouin zone. To understand the phonon-induced blue shift of the magnon frequency, we systematically calculate the exchange interaction as a function of the lattice constant of YIG, as shown in Fig.~\ref{fig:change_a}(a). All these $J$'s decrease as the lattice constant increases and $J_{ad}$ is much larger in magnitude than the others. Using the calculated dominant $J_{ad}$, we can determine the Gr{\"u}neisen constant at the equilibrium lattice constant ($a_0$), $\gamma_m=\partial{\ln}J_{ad}/\partial{\ln}V\vert_{V=a^3_0}=3.07$. It agrees well with experimental values 3.13 and 3.26 (Ref.~\onlinecite{Bloch:jpcs66,Kamilov:pu98}). The calculated magnon spectra, as plotted in Fig.~\ref{fig:change_a}(b), show a monotonic decrease in frequency as the lattice constant increases (along the direction of the arrow). 

Note that the dependence of $J$'s on the lattice constant is not perfectly linear. A detailed inspection shows that the derivative $\partial J/\partial a$ decreases as $a$ increases. Therefore, compressing the lattice leads to a larger rise in the exchange energy than the energy reduction by expanding the lattice by the same amount. This is also reflected by the magnon spectra in Fig.~\ref{fig:change_a}(b), where the magnon frequency differences are not equal but becomes larger at small lattice constants. On the other hand, the lattice vibration can be described by a harmonic potential in the lowest order approximation, where the atom displacement subject to a Gaussian distribution is symmetric with respect to its equilibrium position. As a consequence, the displacements that decrease interatomic distance lead to the rise in the magnon frequency, which is larger than the energy reduction caused by the increase of interatomic distance. This explains the slight blue shift of the magnon spectra at room temperature. 

\begin{figure}
\begin{center} 
\includegraphics[width=\columnwidth]{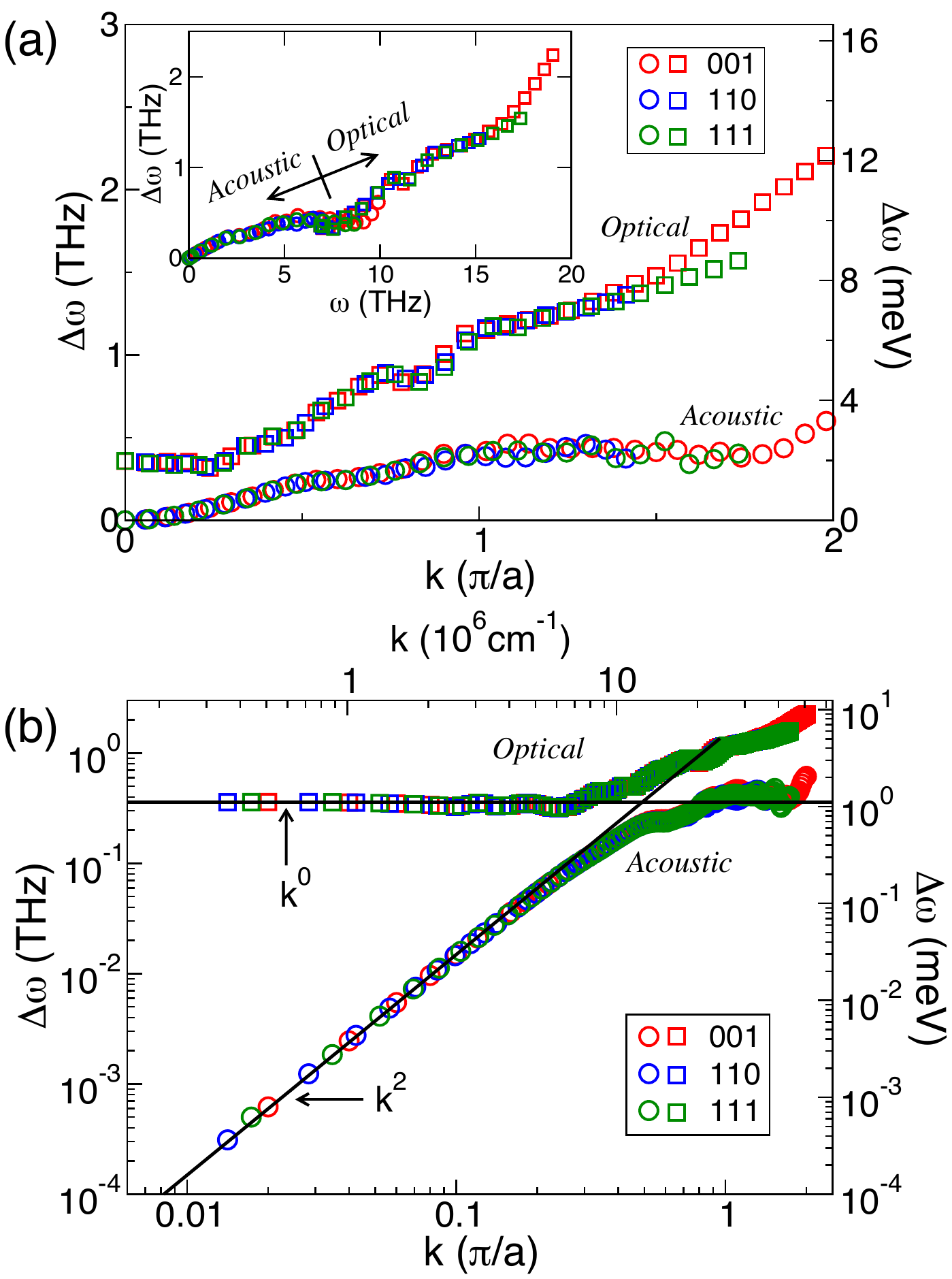}
\end{center}
\caption{(a) Calculated broadening of the magnon spectrum of YIG, $\Delta\omega$, at room temperature as a function of $k$. Inset: $\Delta\omega$ replotted as a function of magnon frequency $\omega$. (b) $\Delta\omega(\mathbf k)$ replotted in a log-log scale. The black lines indicate a constant $\Delta\omega$ and a quadratic dependence on $k$ for the optical branch and the acoustic branch, respectively.}\label{fig:spread}
\end{figure}
Though phonons do not dramatically influence the magnon dispersion, they give rise to a significant broadening of the spectra indicating a finite lifetime of the magnon due to the magnon-phonon interaction. We extract the broadening of the spectra $\Delta\omega$ for the acoustic and the lowest-energy optical magnon modes out of the 40 room-temperature configurations. Figure~\ref{fig:spread}(a) shows the calculated $\Delta\omega$ for both acoustic and optical magnons, both of which increase monotonically with $\vert\mathbf k\vert$ and exhibit very little anisotropy. We further replot $\Delta\omega$ as a function of $\omega$ in the inset of Fig.~\ref{fig:spread}(a), where all the data points fall into a single continuous curve indicating that the magnon-phonon relaxation time only depends on the magnon energy. This may not be surprising since the broadening of magnon excitation results from the phonon-induced fluctuation of the exchange interaction, which in turn only depends on the energy.~\cite{Gunnarsson:jpf76} In addition, as the frequency increases, the density of states of phonons increases monotonically in this frequency range resulting in an increasing magnon-phonon scattering rate. This is the reason why $\Delta\omega$ increases monotonically with the frequency. In Fig.~\ref{fig:spread}(b), $\Delta\omega$ is replotted in the logarithmic scale to see the asymptotic behavior in the long wavelength limit, where we can find a quadratic dependence on $k$ of $\Delta\omega$ for the acoustic branch and a constant $\Delta\omega=1.48$~meV for the optical branch, as illustrated by the black solid lines.

The magnon-phonon relaxation time $\tau_{\textrm mp}$ can be estimated from the broadening $\Delta\omega$ of the magnon spectrum using the uncertainty principle, i.e. $\Delta\omega\tau_{\textrm mp}\approx\hbar$. For the acoustic magnon at small $k$, we have $\tau_{\textrm mp}\propto k^{-2}$ up to about $k=\pi/2a$. Note that this relation qualitatively agrees with the phonon-induced absorption rate of sound waves in solids, \cite{Landau:70} $\tau^{-1}\propto\omega^2=(ck)^2$, where $c$ is the velocity of the sound wave. The latter was used to estimate the magnon relaxation rate \cite{Vittoria:prb10}. If we choose a specific wave vector, for instance, from Ref.~\onlinecite{Agrawal:prl13}, $k=5.67\times10^5$~cm$^{-1}$, the magnon-phonon relaxation time can be estimated as $\tau_{\textrm mp}=0.2$~ns. It is worth mentioning that the quadratic dependence of the relaxation time $\tau_{\textrm mp}$ on the wavevector $k$ may not be extrapolated to a much smaller $k$ in the dipolar magnon regime,  where the magnon frequency is dominated by magnetic dipole-dipole interaction that is not included in our calculation. For the lowest-energy optical magnon, the minimum broadening at small $k$ is a constant, $\Delta\omega=1.48$~meV, corresponding to the magnon-phonon relaxation time $\tau_{\textrm mp}=4.4\times10^{-13}$~s. This is rather small because the long-wavelength optical magnons have the relatively high frequency corresponding to a large density of state of phonons in the same frequency range. In this case, the magnon-phonon scattering rate becomes quite large.

It is interesting to note that a higher order dependence, $k^4$, of the magnon-phonon scattering rate, would be obtained if one employs the oversimplified Heisenberg model of a simple cubic ferromagnet to describe phonon-mediated exchange interaction in YIG, i.e.
\begin{equation}
\mathcal{H}_{mp}=-\frac{1}{2}\sum_{i,j}J(|\mathbf{r}_i-\mathbf{r}_j|)\mathbf{S}_{i}\cdot\mathbf{S}_{j}\,.
\end{equation}
Here phonons contribute to the change of exchange coupling through atomic displacements $\mathbf u_i=\mathbf r_i-\mathbf R_i$ from the equilibrium positions $\mathbf R_i$,
\begin{equation}
J(|\mathbf{r}_i-\mathbf{r}_j|)=J(|\mathbf{R}_i-\mathbf{R}_j|)+\nabla J\cdot (|\mathbf{u}_i-\mathbf{u}_j|)+...\,.
\end{equation}
By expressing the displacements in terms of phonon eigen modes~\cite{Liu:prb15} and the spins in terms of magnons, one can determine the rate for a magnon of $\mathbf k$ scattered to $\mathbf k'$ by a phonon $\mathbf q=\mathbf k'-\mathbf k$, which has a $k^4$ dependence. \cite{Simon:unpublished} However, the oversimplified model does not include the optical phonons that are populated at room temperature, or the multiple magnetic Fe atoms in a unit cell that may introduce the internal degrees of freedom for the relaxation. Instead, our calculation take the material-specific electronic and magnetic structures of YIG into account as well as the temperature-induced ionic vibrations and is therefore more realistic. To formulate the lower order $k^2$ dependence of the magnon-phonon scattering rate that we obtained in our calculations calls for further work. 

The mechanism of magnon relaxation at finite temperature under the current study is the phonon-induced fluctuation of the exchange interaction, which is much larger in energy than the spin-orbit interaction. The latter is not included in our calculation due to the fact that both the calculated magnetic moments and the exchange interaction are hardly influenced by the spin-orbit interaction. In contrary to magnetostriction resulting from the magnetoelastic interaction, whose origin is magnetic anisotropy, as reviewed in Ref.~\onlinecite{Kittel:rmp53}, our results suggest that the dominant effect of phonon upon magnon relaxation arises from the modified exchange interaction by lattice vibration. We would also like to emphasize that the Gilbert damping of YIG arises partly from the magnon-phonon relaxation mechanism and partly from the magnon-magnon relaxation~\cite{Barker:prl16}. The latter is beyond the scope of the present work.

In conclusion, we have investigated the magnon-phonon relaxation in YIG at finite temperature by calculating its magnon spectrum with frozen thermal lattice disorder from first principles. The ionic vibrations in the Debye model is employed to model the phonons in YIG at room temperature. The fluctuation of the exchange interaction between magnetic Fe atoms in YIG is found to be the main mechanism of magnon-phonon interaction. The magnon frequencies are slightly blue shifted by the phonons associated with a significant broadening in the magnon spectra. The latter is used to extract the magnon-phonon relaxation time $\tau_{\textrm mp}$. At small $k$, $\tau_{\textrm mp}$ for the acoustic magnon is proportional to $k^{-2}$ while that for the lowest-energy optical magnon is nearly a constant.

The authors would like to thank Simon Streib for sharing their model results and technical suggestions, and thank Ka Shen for helpful discussions. This work was partly supported by the National Natural Science Foundation of China (Grants No. 61604013) and the Fundamental Research Funds for the Central Universities (Grants No. 2016NT10).

\end{document}